\documentclass[10pt,preprint]{aastex}

\def\be{\begin{equation}}
\def\ee{\end{equation}}
\def\d{{\rm d}}

\def\bh{_{\rm BH}}
\def\bhpr{_{{\rm BH},0}}
\def\bhti{_{{\rm BH},i}}
\def\bhtI{_{{\rm BH,I}}}
\def\bol{_{\rm bol}}

\def\D{_{\rm D}}

\def\Edd{_{\rm Edd}}
\def\I{_{\rm I}}

\def\life{_{\rm life}}

\def\scatter{{\Delta_{\log M\bhpr}}}
\def\sfp{{\sf p}}
\def\Sp{_{\rm Sp}}

\def\dex{{\rm\,dex}}

\def\kms{{\rm\,km\,s^{-1}}}

\def\msun{{\rm\,M_\odot}}

\def\yr{{\rm\,yr}}

\begin{document}
\title{The black hole mass versus velocity dispersion relation
in QSOs/Active Galactic Nuclei: observational appearance and black
hole growth}
\author{Qingjuan Yu\footnotemark[1]$^{~,2,3}$ and Youjun Lu$^{2,3,4}$}
\affil{$^2$Astronomy Department, University of
California at Berkeley, Berkeley, CA 94720\\
$^3$Canadian Institute for Theoretical Astrophysics, 60 St.\ George
Street, Toronto, ON M5S 3H8, Canada. \\
$^4$ Center for Astrophysics,
University of Science and Technology of China, 96 Jinzhai Road, Hefei,
Anhui 230026, People's Repulic of China.
}
\footnotetext[1]{Hubble Fellow.}
\email{yqj,lyj@astro.berkeley.edu}
\begin{abstract}
\noindent
Studies of massive black holes (BHs) in nearby galactic centers have revealed a tight
correlation between BH mass and galactic velocity dispersion.
In this paper we investigate how the BH mass versus velocity dispersion
relation and the nuclear luminosity versus velocity dispersion relation in QSOs/active 
galactic nuclei (AGNs)
are connected with the BH mass versus velocity dispersion relation in local galaxies,
through the
nuclear luminosity evolution of individual QSOs/AGNs and the mass growth of individual
BHs. In the study we ignore the effects of BH mergers and assume that the velocity
dispersion does not change significantly during and after the nuclear activity phase. 
Using the observed correlation in local galaxies and
an assumed form of the QSO/AGN luminosity evolution and BH growth,
we obtain the simulated observational appearance of the BH mass versus velocity
dispersion relation in QSOs/AGNs.
The simulation results illustrate how the BH accretion history  
(e.g., the lifetime of nuclear activity and the possibility that QSOs/AGNs
accrete at a super-Eddington accretion rate at the early evolutionary stage)
can be inferred from the difference between the relation in QSOs/AGNs and that
in local galaxies.
We also show how the difference may be weakened by the flux limit of telescopes.
We expect that a large complete sample of QSOs/AGNs with accurate
BH mass and velocity dispersion measurements will help to quantitatively
constrain QSO/AGN luminosity evolution and BH growth models.
\end{abstract}
\keywords{black hole physics -- galaxies: active -- galaxies: evolution --
galaxies: nuclei -- quasars: general}
\maketitle

\section{Introduction}\label{sec:intro}
\noindent
The existence of massive black holes (BHs) in nearby galactic centers, as 
a prediction of the widely accepted QSO model that QSOs are powered by gas
accretion onto massive BHs (e.g., \citealt{Lynden69,S82,R84}), 
is now believed to be confirmed
\citep[e.g.,][]{KR95,Magorrian98,Gebhardt03,Pinkney03}. Studies of the BHs in
nearby galaxies have revealed that BH mass in nearby galactic centers is
tightly correlated with galactic velocity dispersion
\citep{FM00,Geb00a,Tremaine02} and is also (less tightly) correlated with
the luminosity (or mass) of elliptical galaxies or bulges of spiral/S0 galaxies
(e.g., \citealt{KG01}). These correlations suggest a close link between the
formation and evolution of BHs and their host galaxies. 
However, why these correlations exist and whether they also exist in distant
galaxies have not yet had definite answers.

The purpose of this paper is to investigate the relationship between BH mass
and velocity dispersion in QSOs/active galactic nuclei (AGNs). It is important to study this 
in QSOs/AGNs at least for the following two reasons:
\begin{enumerate}
\item 
Various BH growth models have been proposed to explain the origin of the tight
correlation between BH mass and velocity dispersion in nearby galaxies
\citep[e.g.,][]{SR98,Fabian99,Blandford99,Ostriker00,HK00,BS01,AGR01,ITS03,King03},
including accretion of either baryonic gas or non-baryonic dark matter onto
seed BHs or hierarchical mergers of intermediate-mass BHs (which might be end
products of the first generation of stars, or Population III stars) with masses of
typically a few hundred $M\sun$ \citep[e.g.,][]{Schetal02}. How can
these BH growth models be tested by observations? Among these BH growth
processes, currently BH mergers are unlikely to be observed directly although
detecting gravitational wave signals emitted from BH mergers might be possible
in the future. The possible observational features of accretion of
non-baryonic material are still unclear; only
accretion of baryonic gas, which appears as QSO/AGN phenomena, is detectable
and has been extensively studied. We expect that the investigation of the BH
mass versus velocity dispersion relation in QSOs/AGNs will provide an
observationally achievable way to understand BH growth and the origin of the
correlation of local BHs with their host galaxies.
As a matter of fact, some observational investigations on the relation in
QSOs/AGNs have been performed in the past several years
\citep[e.g.,][]{laor98,wandel99,Geb00b,Ferrarese01,Shields03}.

\item 
Furthermore, according to current observations, the total local BH mass
density is consistent with the total mass density accreted onto BHs during
QSO/AGN phases, which suggests that BH mass growth comes mainly from gas
accretion during QSO/AGN phases, rather than from accretion of
non-baryonic material or mergers of intermediate-mass BHs
\citep{YT02,AR02,Fab03}.  Since
QSOs/AGNs represent the population of galaxies housing growing BHs, the study
of the relationship between BH mass and velocity dispersion in QSOs/AGNs can
provide valuable information on the BH growth history, the evolution of
galaxies,
and the origin of the tight correlation between BH mass and velocity dispersion
in local galaxies.
\end{enumerate}

In this paper we show how the BH mass (or nuclear luminosity) versus velocity
dispersion relation in QSOs/AGNs is connected with the BH mass versus velocity
dispersion relation in local galaxies through the nuclear luminosity evolution
of individual QSOs/AGNs and mass growth of individual BHs.
The basic model assumptions are as follows:
\begin{enumerate}
\item BH mergers are assumed to not be important for BH growth.
In principle, the mass growth of a BH may come from both gas accretion due
to QSO/AGN phases and mergers with other BHs.
However, currently, the BH merger process and rate are very uncertain. 
In addition, comparison of the mass density distribution in nearby galaxies
with that accreted due to QSO phases has
shown that BH mergers are not necessarily required at
least for growth of high-mass ($\ga 10^8\msun$) BHs \citep{YT02}.
\item The velocity dispersion of host galaxies is assumed to not change
significantly during and after the nuclear activity phase.  In principle, 
both nuclear luminosity evolution/BH mass growth and the evolution of host
galaxy velocity dispersions contribute to the evolution of the nuclear
luminosity/BH mass versus velocity dispersion relation.
However, ignoring the velocity dispersion evolution may greatly simplify our
analysis and highlight the effects of nuclear luminosity evolution/BH mass
growth.
Comparison of observations with the predictions obtained by ignoring
velocity dispersion evolution might also provide clues
on the effect of velocity dispersion evolution.
In addition, this assumption has also been adopted in many other physical
models to explain the correlation between BH mass and velocity dispersion in
nearby galaxies (e.g., \citealt{SR98,Fabian99,Blandford99,BS01,AGR01,King03}).
\end{enumerate}
The model includes the following three basic components: 
the BH mass distribution of local galaxies at a given host galaxy velocity dispersion,
the curve of the nuclear luminosity evolution/mass growth of individual BHs, 
and the nuclear luminosity/BH mass distribution in QSOs/AGNs
at the given velocity dispersion.
In principle, given any two of them, the last one can be constrained by the
model.
In this paper we use the first two components as model inputs and the
third one as the model output.

The paper is organized as follows: In \S~\ref{sec:m0sigma} we review current
observational results on the BH mass versus velocity dispersion relation in
local galaxies.  In \S~\ref{sec:msigmaqso} we formulate the connection of
the nuclear luminosity/BH mass versus velocity dispersion relation in QSOs/AGNs 
with the BH mass versus velocity dispersion relation in nearby galaxies
through the nuclear luminosity evolution/mass growth of individual BHs. 
Detailed models on the nuclear luminosity evolution/mass growth of
individual BHs are presented in \S~\ref{sec:calLM}. 
Applying the observational results of the local BH
demography (in \S~\ref{sec:m0sigma}) and the assumed QSO luminosity evolution
and BH growth curves (in \S~\ref{sec:calLM}) to the analysis
made in \S~\ref{sec:msigmaqso}, we predict the BH mass versus galactic
velocity dispersion relation in QSOs/AGNs and show the simulation results of its
observational appearance in \S~\ref{sec:results}. 
Discussion on current observational results of the relation in QSOs/AGNs
is given in \S~\ref{sec:dis}.
Finally, conclusions are summarized in \S~\ref{sec:con}.

In \S~\ref{sec:msigmaqso} readers who are primarily interested in the results
of the formulation might want to skip the detailed mathematical manipulations
and move straight to the model predictions given by equations
(\ref{eq:calPLsigma1}), (\ref{eq:calPMsigma}), and (\ref{eq:calPMsigmasimp})
(with the aid of Table~\ref{tab:tab1}).
The QSO luminosity evolution or BH growth is incorporated into those equations
through the lifetime of the nuclear activity $\tau\life$
and a distribution function of the probability
that the progenitor of a local BH with mass $M_{\rm BH,0}$
(where the subscript ``0'' represents the current cosmic time $t_0$)
had a mass $M_{\rm BH}$ or nuclear luminosity $L$
in its nuclear activity history
[i.e., $P(M_{\rm BH}|M_{\rm BH,0})$ or $P(L|M_{\rm BH,0})$ in
Table~\ref{tab:tab1}].
The observational counterparts of the model predictions are given
by equations (\ref{eq:calPLsigmaQSO}) and (\ref{eq:calPMsigmaQSO}).
Comparison of the observations with the model predictions may help to strengthen
the existing constraints or provide new information on
the QSO/AGN luminosity evolution and BH growth.

In this paper we set the Hubble constant to $H_0=100h\kms$, and the
cosmological model used is ($\Omega_{\rm
M}$,$\Omega_{\Lambda}$,$h$)$=(0.3,0.7,0.65)$.

\section{The BH mass versus velocity dispersion relation in nearby normal
galaxies} \label{sec:m0sigma}
\noindent
In nearby galaxies BH mass ($M\bhpr$) is tightly correlated with galactic
velocity dispersion ($\sigma$) \citep{FM00,Geb00a,Tremaine02}.
The logarithm of the BH mass at a given velocity dispersion $\sigma$
has a mean value given by
\be
\langle \log (M\bhpr)|\sigma\rangle
=A+\gamma\log(\sigma/200\kms)
\label{eq:msigma}
\ee
\citep{Tremaine02}, where $M\bhpr$ is in units of $\msun$, $\gamma=4.02\pm0.32$,
$A=8.18\pm0.06$ has been adjusted to our assumed Hubble constant $h=0.65$
(see section 2.2 in \citealt{YT02}), and $\sigma$ is the luminosity-weighted
line-of-sight velocity dispersion within the effective radius of galaxies.
Note that relation (\ref{eq:msigma}) is fitted in
$\log M\bhpr$--$\log\sigma$ space.
We assume that the distribution in $\log M\bhpr$ at a given $\sigma$ is
Gaussian, with intrinsic standard deviation $\scatter$
(which is assumed to be independent of $\sigma$ here),
and can be written as
\be
\sfp(\log M\bhpr|\sigma,t_0)=\frac{1}{\sqrt{2\pi}\scatter}
\exp\left[-\frac{(\log M\bhpr-\langle \log M\bhpr|\sigma\rangle)^2}
{2\Delta^2_{\log M\bhpr}}\right].
\label{eq:probm0tom}
\ee
According to \citet{Tremaine02}, the intrinsic scatter in $\log M\bhpr$
should not be larger than 0.25--$0.3$~\dex. 
In this paper we set $\scatter=0.27\dex$.

According to Bayes's theorem, the velocity dispersion probability distribution 
function (PDF)
of nearby galaxies with BH mass $M\bhpr$, $\sfp(\sigma|M\bhpr,t_0)$, is related 
to \\$\sfp(M\bhpr|\sigma,t_0)=[\ln(10)M\bhpr]^{-1}\sfp(\log M\bhpr|\sigma,t_0)$
by the equation
\be
\sfp(\sigma|M\bhpr,t_0)=\frac{n_\sigma(\sigma,t_0)\sfp(M\bhpr|\sigma,t_0)}
{n_{M\bh}(M\bhpr,t_0)},
\label{eq:bayes}
\ee
where $n_{M\bh}(M\bhpr,t_0)$ is the local BH mass function (BHMF) defined so
that
$n_{M\bh}(M\bhpr,t_0)\d M\bhpr$ represents the number density of local BHs with
mass in the range $M\bhpr\rightarrow M\bhpr+\d M\bhpr$,
$n_\sigma(\sigma,t_0)$ is the velocity dispersion distribution function of local galaxies
with massive BHs, and we have
$n_{M\bh}(M\bhpr,t_0)=\int n_\sigma(\sigma,t_0)\sfp(M\bhpr|\sigma,t_0)\d\sigma$.

Besides $\sfp(M\bhpr|\sigma,t_0)$ above, several other PDFs are also
defined below.
For comparison of their physical meanings, we list all of them in Table 1.

\begin{table}[hp]
\begin{tabular}{lp{11cm}l}
\tableline
\tableline
symbol & physical meaning & reference\\
\tableline
\tableline
$\sfp(M\bhpr|\sigma,t_0)$
& PDF of BH mass in local galaxies with velocity dispersion $\sigma$
& eq.~\ref{eq:probm0tom}\\
$\sfp(\sigma|M\bhpr,t_0)$ 
& PDF of velocity dispersion in local galaxies with central BH mass $M\bhpr$
& eq.~\ref{eq:bayes}
\\
\tableline
$p(L|\sigma,t)$
& PDF of nuclear luminosity in QSOs/AGNs with velocity dispersion $\sigma$ at cosmic
time $t$
& eqs.~\ref{eq:pLsigma}, \ref{eq:pLsigmaQSO}\\
$p(M\bh|\sigma,t)$
& PDF of BH mass in QSOs/AGNs with velocity dispersion $\sigma$ at cosmic time $t$
& eqs.~\ref{eq:pMsigma}, \ref{eq:pMsigmaQSO}\\
\tableline
${\cal P}(L|\sigma)$
& PDF of nuclear luminosity in QSOs/AGNs at all redshifts and with velocity
dispersion $\sigma$
& eqs.~\ref{eq:calPLsigma}, \ref{eq:calPLsigmaQSO}
\\
${\cal P}(\sigma|L)$
& PDF of velocity dispersion in QSOs/AGNs at all redshifts and with nuclear
luminosity $L$
& eqs.~\ref{eq:calPsigmaL0}, \ref{eq:calPsigmaLQSO}
\\
${\cal P}(M\bh|\sigma)$
& PDF of BH mass in QSOs/AGNs at all redshifts and with velocity dispersion $\sigma$
& eqs.~\ref{eq:calPMsigma0}, \ref{eq:calPMsigmaQSO}
\\
\tableline
$P(L|M\bhpr)$
& PDF of nuclear luminosity of the progenitor of a local BH
during its nuclear activity phases
& eq.~\ref{eq:PLM}
\\
$P(M\bh|M\bhpr)$
& PDF of BH mass of the progenitor of a local BH during its nuclear activity phases
& eq.~\ref{eq:PMMf}
\\
\tableline
\tableline
\end{tabular}
\caption{List of PDFs defined in this paper.}
\label{tab:tab1}
\end{table}

\section{The demography of QSOs/AGNs}\label{sec:msigmaqso}
\noindent
In this section we analytically link the nuclear luminosity/BH mass
versus velocity dispersion relation in QSOs/AGNs
to the QSO/AGN luminosity evolution, BH growth, and demography of local BHs.
We investigate the nuclear luminosity versus velocity dispersion relation
in QSOs/AGNs in \S~\ref{sec:Lprob}
and the BH mass versus velocity dispersion relation in QSOs/AGNs in
\S~\ref{sec:Mprob}.
For each relation, we first present 
the model prediction
and then show its observational counterpart.

\subsection{The nuclear luminosity versus velocity dispersion relation in
QSOs/AGNs}
\label{sec:Lprob}
\subsubsection{Model prediction from local BHs and their nuclear luminosity
evolution}\label{sec:modelpred}
\noindent
We define ${\cal N}_L(t_i,M\bhpr,L,t)$ ($t\ge t_i$) so that ${\cal
N}_L(t_i,M\bhpr,L,t)\d t_i\d M\bhpr\d L\d t$ is the comoving number density of
local BHs with the following properties: the nuclear activity due to accretion onto
their seed BHs\footnote{
Here the original mass of seed BHs does not come from gas accretion (which
appears as QSO/AGN phenomena).
Seed BHs could be remnants of Population III stars, products of dynamical
processes in dense star clusters, or primordial BHs formed in the early
universe, etc. (e.g., \citealt{Marel04}).
The seed BH mass could also be due to non-luminous accretion.
If the nuclear activity of a QSO/AGN is triggered by gas accretion recurrently,
only the BH at the time of the first-time triggering ($t_i$)
is taken as the ``seed BH'' of the QSO/AGN.}
was triggered during cosmic time $t_i\rightarrow t_i+\d t_i$,
the nuclei of their host galaxies were active and had luminosity in the range $L\rightarrow L+\d L$ at cosmic
time $t$, and the BHs are quiescent and have mass in the range $M\bhpr\rightarrow M\bhpr+\d
M\bhpr$ at the present time $t_0$.
Thus, the comoving number density of those BHs whose host galaxies have
velocity dispersions in the range $\sigma\rightarrow\sigma+\d\sigma$ 
is ${\cal N}_L(t_i,M\bhpr,L,t)\sfp(\sigma|M\bhpr,t_0)\d t_i\d M\bhpr\d L\d t\d\sigma$
[see $\sfp(\sigma|M\bhpr,t_0)$ in eq.~\ref{eq:bayes}].
We define the PDF of
the nuclear luminosity of these BHs at cosmic time $t$ whose host galaxies
have velocity dispersion $\sigma$ as follows:
\be
p(L|\sigma,t)\equiv\frac{\int_0^t\d t_i\int \d M\bhpr~{\cal N}_L(t_i,M\bhpr,L,t)\sfp(\sigma|M\bhpr,t_0)}
{\int \d L\int_0^t\d t_i\int \d M\bhpr~{\cal N}_L(t_i,M\bhpr,L,t)\sfp(\sigma|M\bhpr,t_0)}.
\label{eq:pLsigma}
\ee
The function ${\cal N}_L(t_i,M\bhpr,L,t)$ is controlled by both the
rate of triggering accretion onto seed BHs and
the luminosity evolution of individual triggered nuclei.
The rate of triggering nuclear activity is usually believed to be related to
the formation and evolution of galaxies and is a function of cosmic time,
which is very uncertain and not easy to predict.
The luminosity evolution of individual triggered nuclei is believed to contain
information on the accretion process in the vicinity of BHs and is a
function of the physical time that the nuclei have spent since the triggering
of the accretion onto seed BHs.
Note that the luminosity evolution discussed here is different from the
evolution of the characteristic luminosity of the QSO {\em population} as a
function of redshift, which increases with increasing redshift at $z\la 2-3$
and  the variation tendency at $z\ga 2-3$ is not yet clear (e.g., see Fig.~6
in \citealt{Boyle00}).
The luminosity evolution is also not the evolution of the comoving number
density of the QSO {\em population} as a function of redshift:
the comoving number density of QSOs brighter than a certain luminosity has
a peak at redshift $z\sim2-3$ and decreases at both higher and lower redshift.
As with ${\cal N}_L(t_i,M\bhpr,L,t)$, the PDF $p(L|\sigma,t)$
is partly related to the triggering history and is not easy to predict.
By integrating both the numerator and the denominator in equation
(\ref{eq:pLsigma}) over the cosmic time $t$, we define a new nuclear luminosity
PDF of the BHs with host galactic velocity dispersion $\sigma$ as
\be
{\cal P}(L|\sigma)\equiv\frac{\int_0^{t_0}\d t\int_0^t\d t_i\int \d M\bhpr~{\cal N}_L(t_i,M\bhpr,L,t)\sfp(\sigma|M\bhpr,t_0)}
{\int \d L\int_0^{t_0}\d t\int_0^t\d t_i\int \d M\bhpr~{\cal N}_L(t_i,M\bhpr,L,t)\sfp(\sigma|M\bhpr,t_0)}.
\label{eq:calPLsigma}
\ee
As is shown below, to obtain the PDF
${\cal P}(L|\sigma)$, the necessary model inputs are only
the luminosity evolution of individual triggered nuclei and the local
$M\bhpr$-$\sigma$ relation
[see also ${\cal P}(M\bh|\sigma)$ similarly defined in eq.~\ref{eq:calPMsigma0} below].

As done in \citet{YL04}, for BHs with the same mass $M\bhpr$ at present,
we assume
that their nuclear luminosity, ${\cal L}(M\bhpr,\tau)$ is a function only of the age of their nuclear
activity $\tau\equiv t-t_i$. 
Then we use ${\cal L}(M\bhpr,\tau)$ to define two functions to describe the
luminosity evolution of individual triggered nuclei.
One is the lifetime of the nuclear activity for a BH with mass
$M\bhpr$ at present, defined by
\begin{eqnarray}
\tau\life(M\bhpr) & = &\int \d \tau \label{eq:inttau} \\
& = &\int \d L \sum_k \frac{1}{\left|\d {\cal L}(M\bhpr,\tau)/\d \tau|_{\tau=\tau_k}\right|}
\label{eq:taulife}
\end{eqnarray}
(see eqs.~13 and 14 in \citealt{YL04}), where the integration in equation
(\ref{eq:inttau}) is over the period that the nucleus was active and
$\tau_k(L,M\bhpr)$ ($k=1,2,...$) in equation (\ref{eq:taulife}) are the roots
of the equation ${\cal L}(M\bhpr,\tau)-L=0$ ($0<\tau<t_0-t_i$).
Given $L$ and $M\bhpr$, the number of the roots $\tau_k$ can be more than 1,
since ${\cal L}(M\bhpr,\tau)$ can be a {\em non-monotonic} function of $\tau$.
The other function is a PDF of the nuclear luminosity, defined by
\be
P(L|M\bhpr)\equiv \frac{1}{\tau\life(M\bhpr)}\sum_k\frac{1}
{\left|\d {\cal L}(M\bhpr,\tau)/\d \tau|_{\tau=\tau_k}\right|}
\label{eq:PLM}
\ee
so that $\tau\life(M\bhpr)P(L|M\bhpr)\d L$ is the time that a BH 
(with mass $M\bhpr$ at present) spent with nuclear luminosity in the range
$L\rightarrow L+\d L$.
According to equation (17) in \citet{YL04}, ${\cal N}_L(t_i,M\bhpr,L,t)$
is connected with the two functions defined above as follows: 
\be
\int_0^{t_0}\d t\int_0^t\d t_i~{\cal N}_L(t_i,M\bhpr,L,t)
=\tau\life(M\bhpr)P(L|M\bhpr)n_{M\bh}(M\bhpr,t_0).
\label{eq:inttticalN}
\ee
Substituting equation (\ref{eq:inttticalN}) into equation (\ref{eq:calPLsigma}),
we have
\begin{eqnarray}
{\cal P}(L|\sigma)
& = & \frac{\int\d M\bhpr~\tau\life(M\bhpr)P(L|M\bhpr)n_{M\bh}(M\bhpr,t_0)\sfp(\sigma|M\bhpr,t_0)}
{\int\d M\bhpr~\tau\life(M\bhpr)n_{M\bh}(M\bhpr,t_0)\sfp(\sigma|M\bhpr,t_0)}
\label{eq:calPLsigma2}\\
& = & \frac{\int\d M\bhpr~\tau\life(M\bhpr)P(L|M\bhpr)\sfp(M\bhpr|\sigma,t_0)}
{\int\d M\bhpr~\tau\life(M\bhpr)\sfp(M\bhpr|\sigma,t_0)},
\label{eq:calPLsigma1}
\end{eqnarray}
where equation (\ref{eq:bayes}) is used. 

\subsubsection{Observational counterparts}\label{sec:obsQSO}
\noindent
The PDFs of $p(L|\sigma,t)$ and ${\cal P}(L|\sigma)$ can
also be obtained directly from observations of QSOs/AGNs.
Assuming that all the local massive BHs have experienced QSO/AGN phases, 
the observed QSOs/AGNs at redshift $z$ may represent the progenitors of local
galaxies and their central BHs at the cosmic time $t(z)$, and
the QSO luminosity function (QSOLF) $\Psi_L(L,t)$ [defined so that 
$\Psi_L(L,t)\d L$ is the comoving number density of QSOs/AGNs with
luminosity in the range $L\rightarrow L+\d L$ at cosmic time $t$] is given by
(see also \citealt{YL04})
\be
\Psi_L(L,t)=\int_0^t\d t_i\int\d M\bhpr~{\cal N}_L(t_i,M\bhpr,L,t).
\label{eq:QSOLF}
\ee
We can also define the QSO/AGN luminosity and velocity dispersion function (QSOLVF)
$\Phi_{LV}(L,\sigma,t)$ so that $\Phi_{LV}(L,\sigma,t)\d L\d \sigma$ is the comoving number
density of QSOs/AGNs with luminosity in the range $L\rightarrow L+\d L$ at cosmic time $t$
and galactic velocity dispersion in the range $\sigma\rightarrow \sigma+\d \sigma$;
and we have
\be
\Psi_{LV}(L,\sigma,t)=\int_0^t\d t_i\int\d M\bhpr~{\cal N}_L(t_i,M\bhpr,L,t)\sfp(\sigma|M\bhpr,t_0)
\label{eq:QSOLVF}
\ee
\be
\Psi_L(L,t)=\int\Phi_{LV}(L,\sigma,t)\d \sigma.
\label{eq:QSOLF1}
\ee
Substituting equation (\ref{eq:QSOLVF}) into equations (\ref{eq:pLsigma}) and
(\ref{eq:calPLsigma}), we have
\be
p(L|\sigma,t)=\frac{\Phi_{LV}(L,\sigma,t)}{\int\d L~\Phi_{LV}(L,\sigma,t)}
\label{eq:pLsigmaQSO}
\ee
\be
{\cal P}(L|\sigma)=\frac{\int_0^{t_0}\d t~\Phi_{LV}(L,\sigma,t)}
{\int\d L\int_0^{t_0}\d t~\Phi_{LV}(L,\sigma,t)}.
\label{eq:calPLsigmaQSO}
\ee

\subsubsection{${\cal P}(\sigma|L)$}
\noindent
Similar to the definition of ${\cal P}(L|\sigma)$, we may also define a velocity
dispersion PDF in QSOs/AGNs at a given nuclear
luminosity $L$ as follows:
\begin{eqnarray}
{\cal P}(\sigma|L)
& \equiv & \frac{\int_0^{t_0}\d t\int_0^t\d t_i\int \d M\bhpr~{\cal N}_L(t_i,M\bhpr,L,t)\sfp(\sigma|M\bhpr,t_0)}
{\int \d\sigma \int_0^{t_0}\d t\int_0^t\d t_i\int \d M\bhpr~{\cal N}_L(t_i,M\bhpr,L,t)\sfp(\sigma|M\bhpr,t_0)}
\label{eq:calPsigmaL0}\\
& = &\frac{\int\d M\bhpr~\tau\life(M\bhpr)P(L|M\bhpr)n_{M\bh}(M\bhpr,t_0)\sfp(\sigma|M\bhpr,t_0)}
{\int\d M\bhpr~\tau\life(M\bhpr)P(L|M\bhpr)n_{M\bh}(M\bhpr,t_0)}.
\label{eq:calPsigmaL}
\end{eqnarray}
Substituting equations (\ref{eq:QSOLVF}) and (\ref{eq:QSOLF1}) into equation
(\ref{eq:calPsigmaL}), we have the observational counterpart of
${\cal P}(\sigma|L)$ given by
\be
{\cal P}(\sigma|L)=\frac{\int_0^{t_0}\d t~\Phi_{LV}(L,\sigma,t)}
{\int_0^{t_0}\d t~\Psi_L(L,t)}
\label{eq:calPsigmaLQSO}
\ee
(see also eq.~64 in \citealt{YL04}).

As a summary of \S~\ref{sec:Lprob},
equation (\ref{eq:calPLsigmaQSO}) or (\ref{eq:calPsigmaLQSO}) gives the
nuclear luminosity versus velocity dispersion relation in QSOs/AGNs 
directly obtained from observations.
Equation (\ref{eq:calPLsigma1}) or (\ref{eq:calPsigmaL}) shows the model
prediction from the local $M\bhpr$-$\sigma$ relation and 
the luminosity evolution of individual QSOs/AGNs.

\subsection{The BH mass versus velocity dispersion relation in QSOs/AGNs}
\label{sec:Mprob}
\noindent
Given the luminosity evolution of an individual QSO/AGN
${\cal L}\bol(M\bhpr,t)$ 
(where the subscript ``bol'' represents the bolometric luminosity) and
the mass-to-energy conversion efficiency $\epsilon$, the BH mass in QSOs/AGNs
follows the evolution below:
\be
{\cal M}\bh(M\bhpr,\tau)
=M\bhti(M\bhpr)+\int_0^\tau\d\tau\frac{(1-\epsilon){\cal L}\bol(M\bhpr,\tau)}{\epsilon c^2},
\label{eq:calMi}
\ee
or 
\be
{\cal M}\bh(M\bhpr,\tau)
=M\bhpr-\int_\tau^{\tau\life(M\bhpr)}\d\tau\frac{(1-\epsilon){\cal L}\bol(M\bhpr,\tau)}{\epsilon c^2},
\label{eq:mbhattau}
\ee
where $\epsilon$ is assumed to be a constant and $M\bhti$ is the seed BH mass.

As done in \S~\ref{sec:Lprob}, we can also define the BH mass PDF
by replacing $L$ and ${\cal L}(M\bhpr,\tau)$ with $M\bh$
and ${\cal M}\bh(M\bhpr,\tau)$. 
First we define ${\cal N}_{M\bh}(t_i,M\bhpr,M\bh,t)$ ($t\ge t_i$)
so that \\${\cal N}_{M\bh}(t_i,M\bhpr,M\bh,t)\d t_i\d M\bhpr\d M\bh\d t$ is the
comoving number density of local BHs with the following properties: the nuclear activity
due to accretion onto their seed BHs was triggered during cosmic time
$t_i\rightarrow t_i+\d t_i$, the nuclei of their host galaxies were active with
central BH mass in the range $M\bh\rightarrow M\bh+\d M\bh$ at cosmic time
$t$, and these BHs have mass in the range $M\bhpr\rightarrow M\bhpr+\d M\bhpr$
at present time $t_0$.
Similar to equations (\ref{eq:pLsigma}) and
(\ref{eq:calPLsigma}), we can then use ${\cal N}_{M\bh}(t_i,M\bhpr,M\bh,t)$ to
define the BH mass PDF in the nuclear-active progenitors of local BHs,
for example,
\be
p(M\bh|\sigma,t)\equiv
\frac{\int_0^t\d t_i\int \d M\bhpr~{\cal N}_{M\bh}(t_i,M\bhpr,M\bh,t)\sfp(\sigma|M\bhpr,t_0)}
{\int \d M\bh\int_0^t\d t_i\int \d M\bhpr~{\cal N}_{M\bh}(t_i,M\bhpr,M\bh,t)\sfp(\sigma|M\bhpr,t_0)}
\label{eq:pMsigma}
\ee
\begin{eqnarray}
{\cal P}(M\bh|\sigma) &\equiv&
\frac{\int_0^{t_0}\d t\int_0^t\d t_i\int \d M\bhpr~{\cal N}_{M\bh}(t_i,M\bhpr,M\bh,t)\sfp(\sigma|M\bhpr,t_0)}
{\int \d M\bh\int_0^{t_0}\d t\int_0^t\d t_i\int \d M\bhpr~{\cal N}_{M\bh}(t_i,M\bhpr,M\bh,t)\sfp(\sigma|M\bhpr,t_0)}
\label{eq:calPMsigma0}\\
&=&\frac{\int\d M\bhpr~\tau\life(M\bhpr)P(M\bh|M\bhpr)\sfp(M\bhpr|\sigma,t_0)}
{\int\d M\bhpr~\tau\life(M\bhpr)\sfp(M\bhpr|\sigma,t_0)},
\label{eq:calPMsigma}
\end{eqnarray}
where
\be
P(M\bh|M\bhpr)\d M\bh\equiv \frac{\d M\bh}{\tau\life(M\bhpr)}
\frac{1}{\left|\d {\cal M}\bh(M\bh,\tau)/\d \tau\right|_{\tau=\tau_1}},
\label{eq:PMMf}
\ee
and $\tau_1$ is the only root of the equation ${\cal M}\bh(M\bhpr,\tau)-M\bh=0$
($0<\tau<t_0-t_i$) [note that ${\cal M}\bh(M\bhpr,\tau)$ in
eq.~\ref{eq:calMi} or \ref{eq:mbhattau} is a monotonically increasing function
of $\tau$].
The BH growth history (or $\tau\life$) may vary for different $M\bhpr$.
For simplicity, we assume that $\tau\life$ is independent of $M\bhpr$ below,
and thus equation (\ref{eq:calPMsigma}) becomes
\be
{\cal P}(M\bh|\sigma)=\int P(M\bh|M\bhpr)\sfp(M\bhpr|\sigma,t_0) \d M\bhpr.
\label{eq:calPMsigmasimp}
\ee

We can use moments to characterize the distribution of
${\cal P}(\log M\bh|\sigma)=\ln(10) M\bh{\cal P}(M\bh|\sigma)$,
such as the mean of $\log M\bh$ at a given $\sigma$, defined by
\be
\langle\log M\bh|\sigma \rangle\equiv\int
(\log M\bh){\cal P}(\log M\bh|\sigma)\d\log M\bh,
\label{eq:mean}
\ee
and also the standard variance, the skewness, etc.
The difference between the distributions or moments of ${\cal P}(\log M\bh|\sigma)$
and $\sfp(\log M\bhpr|\sigma,t_0)$ contains information on BH accretion history
and/or galaxy evolution.
Obviously, by applying equation  (\ref{eq:calPMsigmasimp}) to equation (\ref{eq:mean}),
we have the difference of the means of the distributions
${\cal P}(M\bh|\sigma)$ and $\sfp(M\bhpr|\sigma,t_0)$,
\be
\delta\langle\log M\bh|\sigma\rangle\equiv\langle\log M\bh|\sigma\rangle-\langle\log M\bhpr|\sigma\rangle<0
\label{eq:meandiff}
\ee 
[note that $P(M\bh|M\bhpr)=0$ if $M\bh>M\bhpr$, and see $\langle\log M\bhpr|\sigma\rangle$
in eq.~\ref{eq:msigma}].
A positive difference of the means might suggest that
galaxy velocity dispersions should increase during or after the nuclear activity
(i.e., the assumption that the velocity dispersion does not significantly
change during or after the nuclear activity should be revised).

Similar to defining the QSOLVF as $\Phi_{LV}(M\bh,\sigma,t)$ in \S~\ref{sec:Lprob}, 
we can define the BH mass and velocity dispersion function in QSOs/AGNs (QSOMVF) as
$\Phi_{MV}(M\bh,\sigma,t)$, so that $\Phi_{MV}(M\bh,\sigma,t)\d M\bh\d \sigma$ 
represents the comoving number density of QSOs/AGNs with central BH mass in
the range $M\bh\rightarrow M\bh+\d M\bh$ and host galactic velocity dispersion
in the range $\sigma+\d\sigma$ at cosmic time $t$. 
Thus we have
\be
p(M\bh|\sigma,t)=\frac{\Phi_{MV}(M\bh,\sigma,t)}{\int\d M\bh~\Phi_{MV}(M\bh,\sigma,t)}
\label{eq:pMsigmaQSO}
\ee
\be
{\cal P}(M\bh|\sigma)=\frac{\int_0^{t_0}\d t~\Phi_{MV}(M\bh,\sigma,t)}
{\int\d M\bh\int_0^{t_0}\d t~\Phi_{MV}(M\bh,\sigma,t)}.
\label{eq:calPMsigmaQSO}
\ee
The distributions of $p(M\bh|\sigma,t)$ and ${\cal P}(M\bh|\sigma)$ give
the BH mass versus velocity dispersion relations for QSOs/AGNs at a given
cosmic time and for QSOs/AGNs at all redshifts, respectively. 
The ${\cal P}(M\bh|\sigma)$ is the $M\bh$-$\sigma$ relation in QSOs/AGNs
that is mainly discussed in this paper.

Comparison of the expectation from equation (\ref{eq:calPMsigmasimp}) (or
[\ref{eq:calPMsigma}], [\ref{eq:calPLsigma1}], [\ref{eq:calPsigmaL}]) with
observations of QSOs/AGNs (see eq.~[\ref{eq:calPMsigmaQSO}], [\ref{eq:calPLsigmaQSO}],
or [\ref{eq:calPsigmaLQSO}]) can provide feedback to our understanding of the
luminosity evolution of individual QSOs/AGNs, the mass growth of individual BHs,
and the evolution of host galaxy velocity dispersions.

\section{The QSO/AGN luminosity evolution and BH growth model:
${\cal L}\bol(M\bhpr,\tau)$ and ${\cal M}\bh(M\bhpr,\tau)$}
\label{sec:calLM}
\noindent
In this section we present the detailed form of the luminosity
evolution and BH growth [${\cal L}\bol(M\bhpr,\tau)$ and
${\cal M}\bh(M\bhpr,\tau)$] incorporated in the analysis
in \S~\ref{sec:msigmaqso}.
We assume that the luminosity evolution of individual QSOs/AGNs can be
described by two phases (see also \S~2.4 in \citealt{YL04}): (1) after the
accretion onto a seed BH is triggered, initially there is sufficient material
to feed the BH, and the BH accretes with Eddington luminosity; (2) with
the mass growth of the BHs and the consumption of the material to feed it,
the material becomes insufficient to support Eddington accretion, and then
the nuclear luminosity declines and is fainter than the Eddington luminosity
(see also \citealt{YL04}).  For simplicity, below we assume that the two phases
appear only once for each BH (see also \citealt{YL04}).

We assume that the first phase lasts for a period of $\tau\I$,
and the BH mass increases to $M\bhtI$ at time $t=t_i+\tau\I\equiv t\I$.
Thus, the nuclear luminosity in the first phase increases with time as
\be
{\cal L}\bol(\tau)=L\Edd(M\bhtI)
\exp\left(\frac{\tau-\tau\I}{\tau\Sp}\right) 
\qquad 0<\tau<\tau\I,
\label{eq:Lphase1}
\ee
where $L\Edd(M\bhtI)$ is the Eddington luminosity of a BH with mass $M\bhtI$, and
\be
\tau\Sp=4.5\times 10^7 \left[\frac{\epsilon}{0.1(1-\epsilon)}\right]\yr
\label{eq:tauSp}
\ee
is the Salpeter time (the time required for a BH radiating at Eddington luminosity
to $e$-fold in mass).
The BH mass increases as
\be
{\cal M}\bh(\tau)=M\bhtI
\exp\left(\frac{\tau-\tau\I}{\tau\Sp}\right)
\qquad 0<\tau<\tau\I.
\label{eq:Mphase1}
\ee
In the second phase we assume that the evolution of the nuclear luminosity
declines as
\be
{\cal L}\bol(\tau)=\cases{L\Edd(M\bhtI)\exp\left(-\frac{\tau-\tau\I}{\tau\D}\right), & for $\tau\I\le\tau\le\tau\I+\xi\tau\D$, \cr
0, & for $\tau>\tau\I+\xi\tau\D$,}
\label{eq:Lphase2}
\ee
where $\tau\D$ is the characteristic declining timescale of the nuclear
luminosity.
We assume that QSOs/AGNs become quiescent when the nuclear luminosity declines
by
a factor of $\eta=\exp(-\xi)$ compared to the peak luminosity $L\Edd(M\bhtI)$,
so there is a cutoff of the nuclear luminosity at
$\tau=\tau\I+\xi\tau\D$ in equation (\ref{eq:Lphase2}).
The factor $\xi$ is set to $\ln(10^3)=6.9$ (as in \citealt{YL04}).
According to equations (\ref{eq:calMi}) and (34), the BH mass increases as
\be
{\cal M}\bh(\tau)=
M\bhtI\left\{1+\frac{\tau\D}{\tau\Sp}\left[1-\exp\left(\frac{\tau\I-\tau}{\tau\D}\right)\right]\right\}, 
\qquad \tau\I\le\tau\le\tau\I+\xi\tau\D.
\label{eq:Mphase2}
\ee
With the assumption that the nuclear activity of all QSOs/AGNs is quenched
at present
(i.e., $t_0-t_i-\tau\I \gg \tau\D$), the BH mass at present is given by
\be
M\bhpr=\left(1+\frac{\tau\D}{\tau\Sp}\right)M\bhtI.
\label{eq:Mbhpr}
\ee

Some constraints on ${\cal L}\bol(M\bhpr,\tau)$ have been obtained by
comparing the time integral of the QSOLF in observations with the prediction
from the local BHMF in \citet{YL04}.  For example, for the nuclear luminosity 
evolution models above, we
should have $\tau\I\ga \tau\Sp$ if $\epsilon \sim 0.1$ and $\tau\D=0$
and have $\tau_I\ga 0.2\tau\Sp$ if $\epsilon\sim 0.31$ and $\tau\D=0$.
The characteristic declining timescale of the second phase, $\tau\D$, should
be significantly
shorter than $\tau\Sp$, and BH growth should not be dominated by accretion
in the second phase. In this paper, based on those constraints,
we assume two family models for the parameter $\tau\D$: 
(1) $\tau_D=0$ and (2) $\tau_D=0.3\tau\Sp$. In each of the family
models, we consider four cases for the total mass increase of a BH during
its nuclear activity period: that is,
$M\bhpr/M\bhti=\exp(0.3),\exp(1),\exp(2)$, and $\exp(4)$, which
are denoted as cases A-D,
respectively.  Thus, we have $\tau\D=0$ and $\tau\I/\tau\Sp=0.3,~1,~2$, and $4$ for
models 1A-1D; and we have $\tau\D/\tau\Sp=0.3$ and
$\tau\I/\tau\Sp=0.04,~0.7,~1.7$, and $3.7$ for models 2A-2D.

\section{Simulation results of the $M\bh$-$\sigma$ relation in QSOs/AGNs}
\label{sec:results}
\noindent
In this section, by applying the BH growth models in \S~\ref{sec:calLM} to
the analysis in \S~\ref{sec:msigmaqso}, we obtain the expected observational
appearance of the $M\bh$-$\sigma$ relation
in QSOs/AGNs. We quantitatively illustrate how this relation in QSOs/AGNs depends on 
the BH growth history and how the flux limit of telescopes
affects the observational appearance of the relation.
Similarly, the expected observational appearance of the $L$-$\sigma$ relation in QSOs/AGNs
can also be obtained, but for simplicity we do not show and discuss it in this
paper.

\subsection{$\tau\I$ and $\tau\D$}\label{sec:tauID}
\noindent
With the local $M\bhpr$-$\sigma$ relation [or $\sfp(M\bhpr|\sigma,t_0)$; see
eqs.~\ref{eq:msigma} and \ref{eq:probm0tom}] and the QSO/AGN
luminosity evolution and BH growth models described in \S~\ref{sec:calLM}, we use
equation (\ref{eq:calPMsigmasimp}) to get the distribution ${\cal P}(M\bh|\sigma)$ in
QSOs/AGNs. The results of ${\cal P}(\log M\bh|\sigma=200\kms)$ are shown in
Figure~\ref{fig:f1}a for model 1 and in Figure~\ref{fig:f1}b for model 2. In each
panel we show the distribution in nearby normal galaxies, $\sfp(\log M\bhpr|\sigma,t_0)$, by the
solid line and
show the distribution in QSOs/AGNs, ${\cal P}(\log M\bh|\sigma)$, by the dotted,
short-dashed, dot-dashed, and long-dashed lines for cases A-D,
respectively.
Note that the shape of the distribution $\sfp(\log M\bhpr|\sigma,t_0)$ has been
assumed to be independent of $\sigma$, and the characteristic increasing and
decreasing timescales ($\tau\I$, $\tau\D$) have
been assumed to be independent of $M\bhpr$; therefore, the shapes of the curves
of ${\cal P}(\log M\bh|\sigma)$ are independent of $\sigma$, and the
distributions for other velocity dispersions can be obtained simply by shifting
the curves to higher or lower BH masses.
The difference of the distributions ${\cal P}(\log M\bh|\sigma)$ and $\sfp(\log M\bhpr|\sigma,t_0)$
can be characterized by the difference of the mean
($\delta\langle \log M\bh|\sigma\rangle$ in eq.~\ref{eq:meandiff}),
the standard variance, the skewness, etc.
For simplicity, we mainly discuss the difference of the mean below.

As seen from Figure~\ref{fig:f1}, with increasing timescale $\tau\I$
(see eq.~\ref{eq:Lphase1}), the curve peak of ${\cal P}(\log M\bh|\sigma)$ shifts
to low BH mass and the scatter of ${\cal P}(\log M\bh|\sigma)$ increases,
since the probability of observing QSOs/AGNs with low BH masses increases.
Our calculations show that we have the difference $\delta \langle\log
M\bh\rangle=\log(M\bhti/M\bhpr)/2\simeq -0.07,-0.2,-0.4$ and $,-0.9$~\dex for models
1A-1D in panel (a) and $\delta \langle\log M\bh\rangle\simeq -0.02,-0.08,-0.2$, and $-0.6$~\dex
for models 2A-2D in Figure~\ref{fig:f1}b.
The difference $\delta\langle\log M\bh\rangle$ is difficult to
detect for cases (A) and (B), considering the
current large measurement errors of BH masses in QSOs/AGNs (see the discussion in
\S~\ref{sec:dis}), and the difference is relatively significant for case D,
which should be detectable if there existed a
sufficiently large and complete sample of QSOs/AGNs.
As seen from Figure~\ref{fig:f1}, for the same case of the two family models,
compared to model 1 ($\tau\D=0$), ${\cal P}(\log M\bh|\sigma)$ in
model 2 ($\tau\D=0.3\tau\Sp$) is lower at the low-mass end and
higher at the high-mass end, and the peak of the curve is closer to the peak
of the local distribution $\sfp(\log M\bhpr|\sigma,t_0)$, although the BH mass is
increased by the same factor during the nuclear activity.
The reason is that in model 2, QSOs/AGNs have a
substantially large probability of being observed in the second phase (see
eq.~\ref{eq:Lphase2}), in which the QSOs/AGNs are shining at sub-Eddington
luminosities and BH masses grow relatively slowly and have acquired nearly all
of their mass.  In addition, Figure 1(b) shows that the existence of
sub-Eddington accretion at the late evolutionary stage may also introduce skewness
to the distribution of ${\cal P}(\log M\bh|\sigma)$.

\subsection{Super-Eddington accretion}\label{sec:supEdd}
\noindent
In Figure~\ref{fig:f1} the nuclear luminosity has been assumed to be
limited by the Eddington luminosity. In the literature, however, it is 
argued that QSOs/AGNs might be able to accrete material at a rate higher than
the
Eddington accretion rate (which is conventionally defined by the Eddington
luminosity for $\epsilon=0.1$ here) at the early stage of their evolution (since
initially the material supply for the BH growth may be sufficiently large;
e.g., \citealt{Blandford03}).  Below we use a simple example to illustrate how
the $M\bh$-$\sigma$ relation in QSOs/AGNs is affected by super-Eddington
accretion.

We modify model 1D ($\tau\I/\tau\Sp=4$) as follows: BHs first accrete
material at a rate higher than the Eddington rate by a factor of $l$
in the
initial $3\tau'\Sp$ period of the nuclear activity, where $l$ is a constant and
$\tau'\Sp=l^{-1}\tau\Sp$ (eq.~\ref{eq:tauSp}), and then BHs accrete material at
the Eddington rate for a period of $\tau\Sp$.  We denote the modified model as
model 1D$'$. 
The BH mass is increased by the same factor 
in the initial $3\tau'\Sp$ period of the nuclear activity in model 1D$'$
as that in the initial $3\tau\Sp$ period of the nuclear activity in model 1D.
The total BH mass increase over the whole nuclear activity period
is also the same in models 1D$'$ and 1D.
The ${\cal P}(\log M\bh|\sigma)$ of model 1D$'$ are shown as
the dotted ($l=2$) and dot-dashed ($l=10$) lines in Figure~\ref{fig:f2}. 
For comparison, $\sfp(\log M\bhpr|\sigma,t_0)$ and the results of models 1B
($\tau\I/\tau\Sp=1$) and 1D shown in Figure~\ref{fig:f1}a are also shown
in Figure~\ref{fig:f2} as the solid, short-dashed, and long-dashed lines, respectively.
As seen from Figure~\ref{fig:f2}, the
dot-dashed line (model 1D$'$ with $l=10$) is lower than both the dotted
(model 1D$'$ with $l=2$) and long-dashed lines (model 1D) at the low-mass end
($\la 10^7\msun$), and the distribution of the dot-dashed line is closer to that of the
short-dashed line (model 1B; $\tau\I/\tau\Sp=1$) than that of the 
dotted line. The reason is that the larger the accretion rate, the more
rapidly the BH mass increases, and thus the smaller the probability
of the {\em low-mass} BHs within a unit mass range being observed, since the
high-rate or super-Eddington accretion is at the early BH growth
stage here. For
example, in model 1D$'$, the BH mass increases by a factor of $\sim 20$ at
the early super-Eddington accretion stage and by a factor of $\sim 2.7$ at the
Eddington accretion stage; for the case of $l=2$, the time fractions for a QSO
to be at those stages are 60\% and 40\%, respectively,
but for the case of $l=10$, the corresponding time fractions become
only 23\% and up to 77\%. The
effect of the super-Eddington accretion can also be quantitatively 
characterized by high-order moments of ${\cal P}(\log M\bh|\sigma)$, such as
skewness.

\subsection{Effects of the flux limit of telescopes}\label{sec:detection}
\begin{figure}
\epsscale{0.95}
\plotone{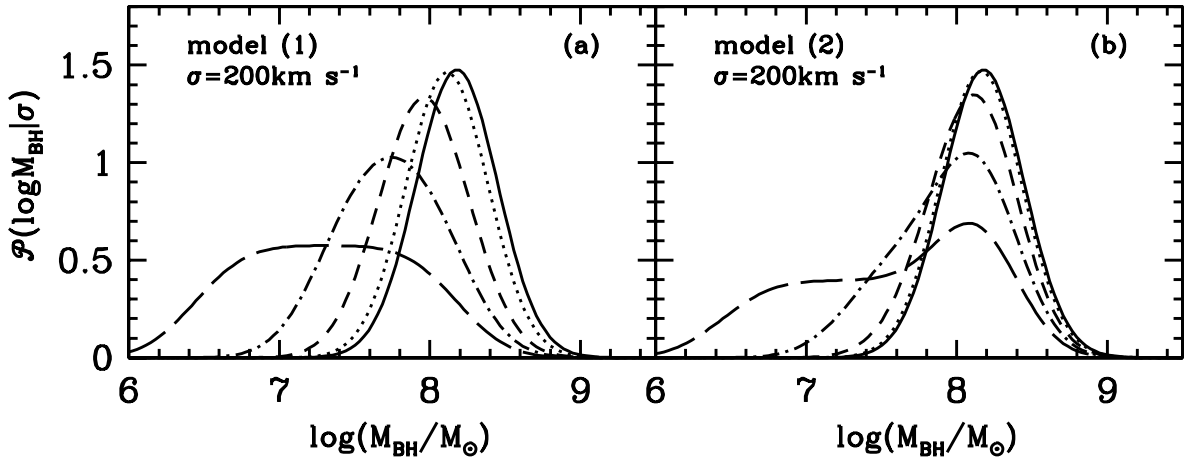}
\caption{Estimated BH mass distribution in QSOs/AGNs [i.e., ${\cal P}(\log
M\bh|\sigma)$; see eq.~\ref{eq:calPMsigmasimp}] at a given host galaxy velocity
dispersion $\sigma=200\kms$. The luminosity evolution and BH growth models used
are described in \S~\ref{sec:calLM}. In (a) are the results for model 1
($\tau\D=0$) and (b) is for model 2 ($\tau\D/\tau\Sp=0.3$).  In both
panels, the solid line represents $\sfp(\log M\bhpr|\sigma,t_0)$ (or the local
$M\bhpr$-$\sigma$ relation, for which a Gaussian distribution of $\log M\bh$ is
assumed with an intrinsic scatter of $0.27$~\dex; see eqs.~\ref{eq:msigma} and
\ref{eq:probm0tom}). The dotted, short-dashed, dot-dashed, and long-dashed
lines represent the results obtained by setting $\ln(M\bhpr/M\bhti)=0.3,1,2$, and $4$
(which are denoted by cases A-D, respectively). That is, we have
$\tau\I/\tau\Sp=0.3,1,2$,and $4$ for models 1A-1D in (a), and have
$\tau\I/\tau\Sp=0.04,0.7,1.7$, and $3.7$ for models 2A-2D in (b). 
As seen from this figure, with
increasing $\tau\I$ the probability of ${\cal P}(\log M\bh|\sigma)$ at the
low-mass end increases, and the deviation of ${\cal P}(\log M\bh|\sigma)$ from
$\sfp(\log M\bhpr|\sigma,t_0)$ increases.  Compared to the results of the
purely Eddington accretion model in (a), the existence of sub-Eddington
accretion at the late evolution stage ($\tau\D\neq 0$ in [b]) decreases the
difference $|\delta\langle\log M\bh\rangle|$ (eq.~\ref{eq:meandiff}) and also
may introduce skewness to the distribution of ${\cal P}(\log M\bh|\sigma)$.
See details in \S~\ref{sec:tauID}.
}
\label{fig:f1}
\end{figure}

\begin{figure}
\epsscale{0.8}
\plotone{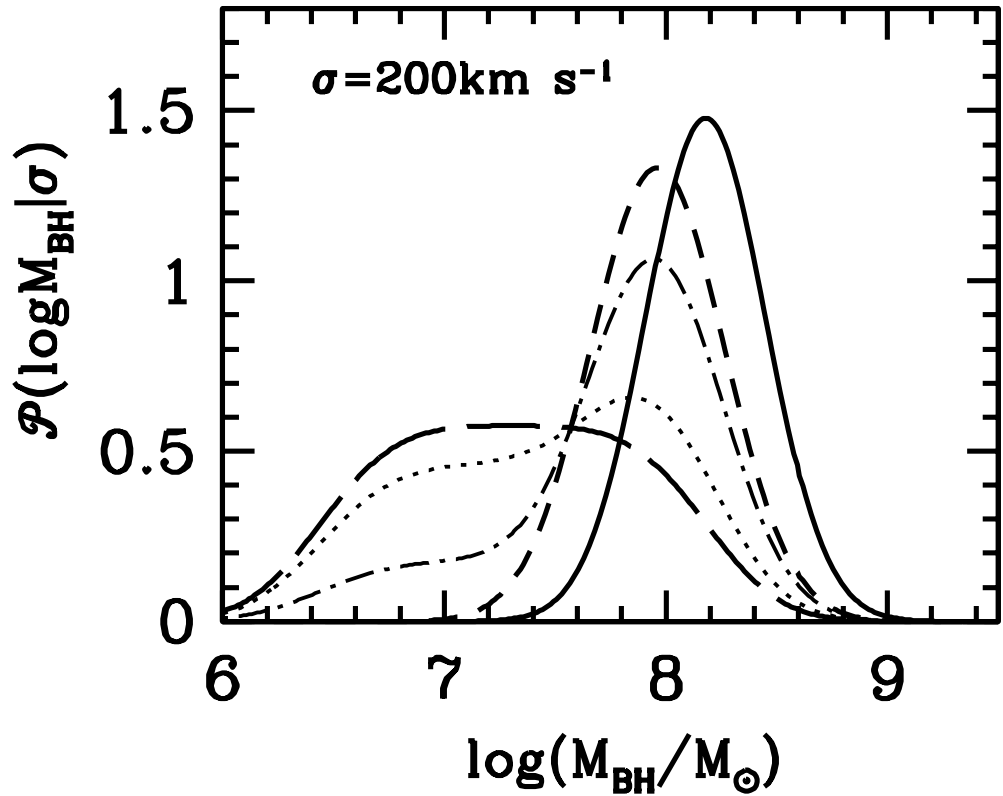}
\caption{Estimated BH mass distribution ${\cal P}(\log M\bh|\sigma)$ 
in QSOs/AGNs at a given host galaxy velocity dispersion, for which
a super-Eddington accretion at the early evolutionary stage of the nuclear
activity is assumed.  The solid,
short-dashed, and long-dashed lines are the same as those in
Figure~\ref{fig:f1}(a).  The dotted and dot-dashed lines represent the results
obtained by model 1D$'$, assuming that BHs accrete material at the
Eddington rate only at the final $\tau\Sp$ period of the nuclear activity.
Before that period, BHs accrete at a rate higher than the Eddington rate,
for example, by a factor of 2 ({\it dotted line}) or 10 ({\it dot-dashed line}). 
As seen from
this figure, super-Eddington accretion at the early stage decreases ${\cal
P}(M\bh|\sigma)$ at the low-mass end and may also introduce skewness
to the distribution ${\cal P}(\log M\bh|\sigma)$.
See details in \S~\ref{sec:supEdd}. }
\label{fig:f2} \end{figure}

\begin{figure}
\epsscale{0.9}
\plotone{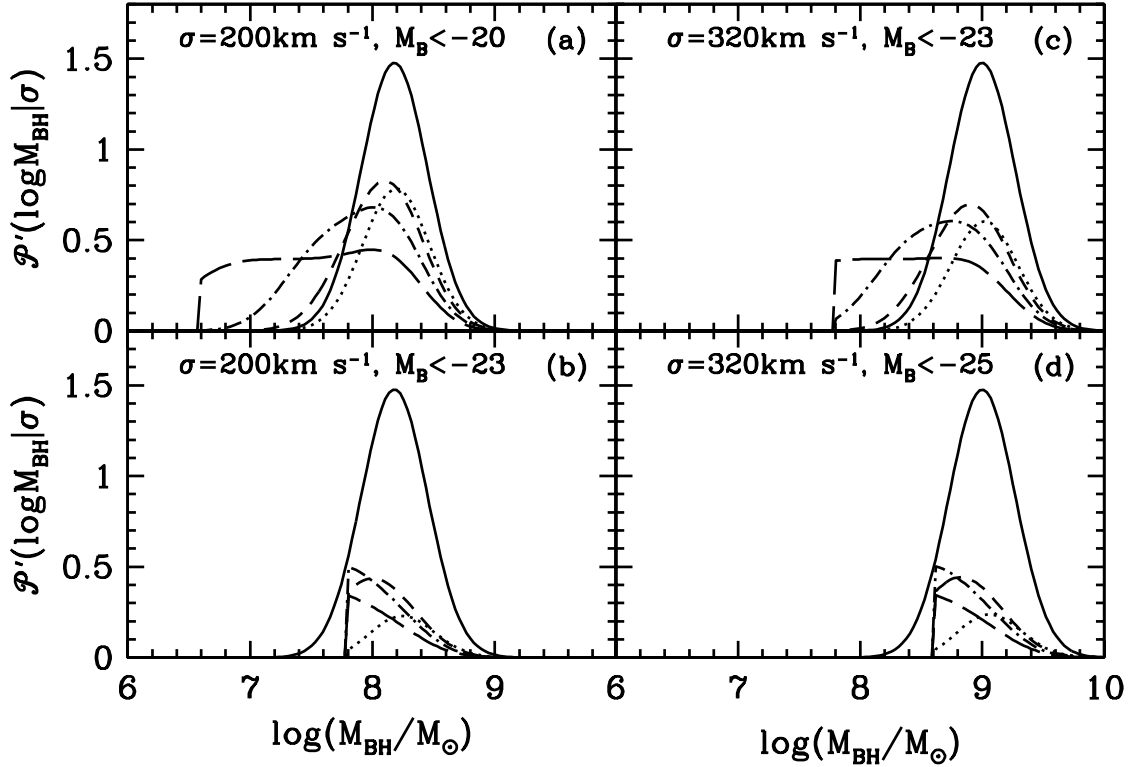}
\caption{Estimated ${\cal P}'(\log M\bh|\sigma)$, i.e., the portion of
${\cal P}(\log M\bh|\sigma)$ contributed to by QSOs/AGNs brighter than a certain
absolute magnitude. 
The line types have the same meanings as in Figure~\ref{fig:f1}.
The $B$-band absolute magnitude truncations are
$M_B<-20,-23$, and $-25$ in (a), (b-c), and (d), respectively.  Other model
parameters ($\tau\I$, $\tau\D$) are the same as in model 2 (see
Fig.~\ref{fig:f1}b).  Panels a and b are for the distributions at
$\sigma=200\kms$, and panels c and d are for the distributions at
$\sigma=320\kms$. 
This figure (and also Fig.~\ref{fig:f4} below) suggests that
the flux limit of telescopes may decrease the difference 
$|\delta\langle\log M\bh\rangle|$ between the observed $M\bh$-$\sigma$ relation in
QSOs/AGNs and the $M\bhpr$-$\sigma$ relation in local galaxies, since
small BHs, the main contributors to the difference,
may be missing from observations.
See details in \S~\ref{sec:detection}.  } \label{fig:f3} \end{figure}

\noindent
In the results above, we have assumed that QSOs/AGNs with any luminosities can
be observed, without being limited by the flux limit of telescopes.  In this
subsection we show how the observational appearance of the $M\bh$-$\sigma$
relation in QSOs/AGNs is affected if only those QSOs/AGNs brighter than a
certain absolute magnitude are detected.  We show the results of ${\cal
P}'(\log M\bh|\sigma)$, the portion of ${\cal P}(\log M\bh|\sigma)$ contributed
to QSOs/AGNs with such absolute magnitude truncations, in Figure~\ref{fig:f3}.
Unlike those of ${\cal P}(\log M\bh|\sigma)$ in Figures~\ref{fig:f1} and \ref{fig:f2},
the shapes of ${\cal P}'(\log M\bh|\sigma)$ vary for different
galactic velocity dispersions because of the magnitude truncation. Thus, we
show the results at two different velocity dispersions: $\sigma=200\kms$ in
Figure 3a and 3b, and $\sigma=320\kms$ in Figure 3c and 3d.
For $\sigma=200\kms$ we set the magnitude truncation of the
QSOs/AGNs to be $M_B<-20$ and $-23$ (the bolometric correction
for the $B$-band luminosity, $C_B$, is defined through
$L\bol\equiv C_B L_{\nu_B}$, where $L_{\nu_B}$ is the energy
radiated at the central frequency of the $B$ band per unit time and
logarithmic interval of frequency, and we set $C_B=11.8$; see \citealt{Elvisetal}).  
For the high-velocity dispersion $\sigma=320\kms$, we set a brighter
magnitude truncation ($M_B<-23$ and $-25$) because of the following considerations:
at $z\la$ 2--3 the characteristic
luminosity of QSOs/AGNs increases with the increase of redshift (see the QSOLF
given in \citealt{Boyle00}), or high-mass BHs are more likely to grow up at high
redshifts, and for high-redshift objects, only the bright ones are observed.
Other model parameters in
Figure~\ref{fig:f3} ($\tau\I$, $\tau\D$) are the same as those in models
2A-2D (see Fig.~\ref{fig:f1}b). As seen from Figure~\ref{fig:f3},
QSOs/AGNs housing small BHs cannot be detected because of the faintness cutoff.
The distribution of ${\cal P}'(\log M\bh|\sigma)$ is also lower
than that of ${\cal P}(\log M\bh|\sigma)$
at the high-mass end because some QSOs/AGNs that house big BHs but
are at the late stage of the second luminosity evolutionary phase
(eq.~\ref{eq:Lphase2}) may also be missing from observations. 
Our calculations show that if the sample of QSOs with $M_B<-20$ is complete,
$|\delta\langle\log M\bh\rangle|$ is about $0.6$~\dex at $\sigma=200\kms$
for model 2D (see the long-dashed line in Fig.~\ref{fig:f3}a), which is
detectable if the accuracy of the BH mass measurement is within a factor of
$\sim 2$-$3$. However, if the magnitude truncation is $M_B<-23$,
we have the difference $|\delta\langle\log M\bh\rangle|\la 0.1\dex$ at
$\sigma=200\kms$ (see Fig.~\ref{fig:f3}b), which is too
small to detect currently.
In short, Figure~\ref{fig:f3} suggests that the flux limit of telescopes may
weaken the
difference of the observed $M\bh$-$\sigma$ relation in QSOs/AGNs from the local
$M\bhpr$-$\sigma$ relation, since small BHs, the main contributors to the
difference of these relations, may be missing from observations.

\begin{figure}
\epsscale{0.8}
\plotone{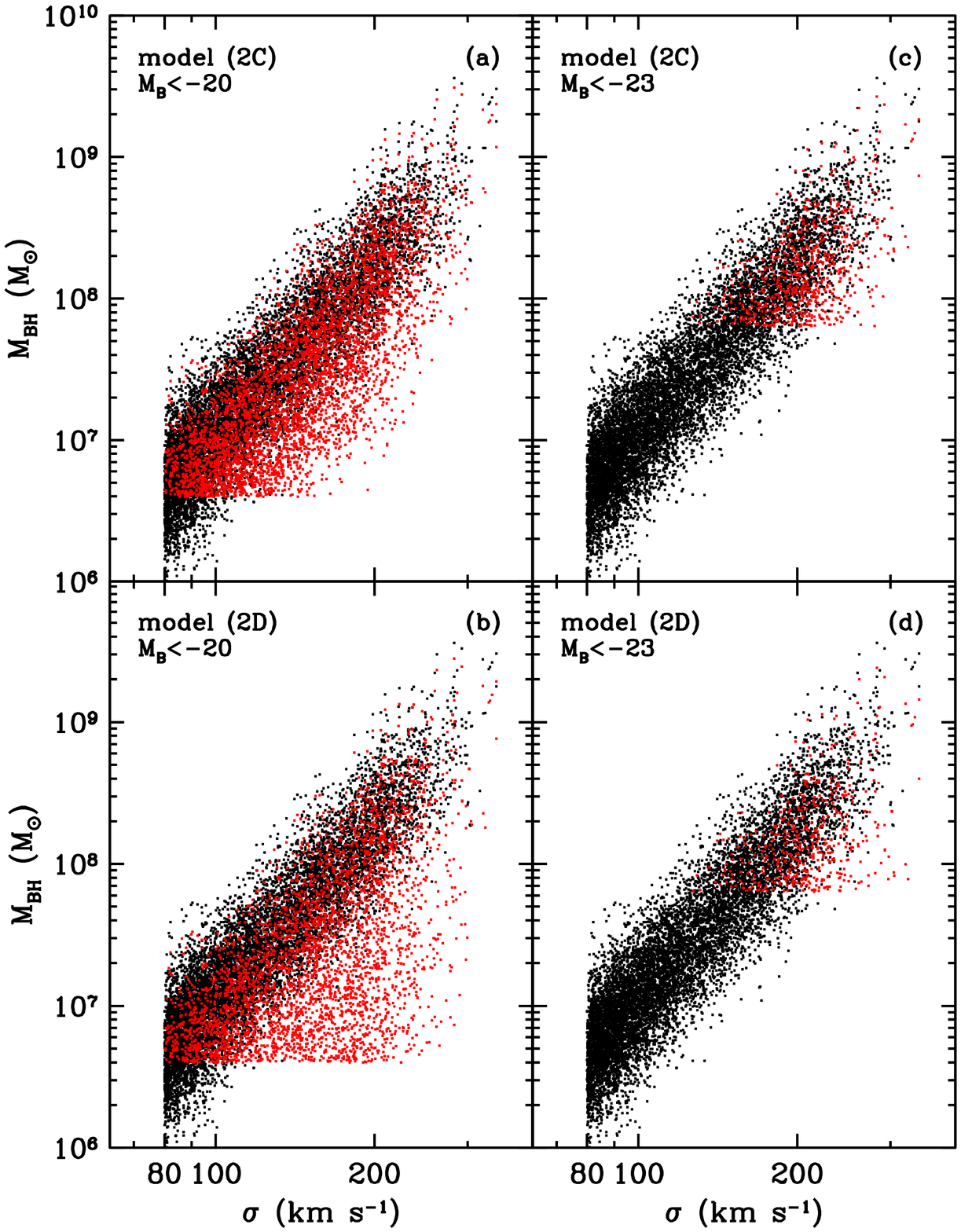}
\caption{Simulation results of the $M\bh$-$\sigma$ relation in QSOs/AGNs.
The black dots represent local galaxies with an assumed intrinsic scatter in
$\log M\bhpr$ of $0.27$~\dex, and the red dots represent
QSOs/AGNs. 
Panels a and c are for model 2C, and panels b and d are for
model 2D.
Panels a and b only include QSOs/AGNs with $M_B<-20$, and panels c and d
only include QSOs/AGNs with $M_B<-23$.
Each panel has $10^4$ black dots; and the numbers of red
dots are 4276, 3562, 658, and 462 in panels (a-d), respectively.
See details in \S~\ref{sec:detection}.
}
\label{fig:f4}
\end{figure}

To further illustrate the appearance of the $M\bh$-$\sigma$ relation in
QSOs/AGNs, we simulate a sample of local galaxies and QSOs/AGNs and show the
result in an $M\bh$ versus $\sigma$ plot (see the black dots
for local galaxies and red dots for QSOs/AGNs in Fig.~\ref{fig:f4}).  To do
this, we first use the Monte-Carlo method to draw a sample of $10^4$ 
galaxies from the velocity dispersion distribution of local galaxies
with $\sigma>80\kms$ (see the black dots in Fig.~\ref{fig:f4}; for the velocity
dispersion distribution of local galaxies, see \citealt{YL04}).
Then, for each (local) galaxy, we use the Monte-Carlo method to generate one
progenitor (i.e., QSOs/AGNs) from the distribution function ${\cal P}(M\bh|\sigma)$.
Note that the number of progenitors generated for each local galaxy, which
should be proportional to the lifetime of the nuclear activity of this
galaxy $\tau\life$, is the same, since the lifetime has been assumed to be the same
for all the galaxies here.
In addition, although only one progenitor is produced for each galaxy,
the total number of progenitors generated for all the galaxies is
large enough in this paper to illustrate the appearance of the $M\bh$-i$\sigma$ relation
in QSOs/AGNs.
Finally, we cut off faint progenitors, and the rest are just the
sample of QSOs/AGNs with magnitude truncation (red dots in Fig.~\ref{fig:f4}).
We show the results of models
2C and 2D ($\tau\D/\tau\Sp=0.3$, $\tau\I/\tau\Sp=2,4$) with an absolute
magnitude truncation of $M_B<-20$ in Figures 4a and 4b, and show the results
of models 2C and 2D with $M_B<-23$ in Figures 4c and 4d, respectively.
The simulation results of models 2A and 2B are not shown in this figure
because they are visually difficult to distinguish from the local
$M\bhpr$-$\sigma$ relation. In models 2C
and 2D, the increase of the BH mass via accretion is large enough (e.g., by a
factor of $\ga 10$) that as seen from Figures~\ref{fig:f4}a and 4b, if all or
at least most of the QSOs/AGNs with absolute magnitude $M_B<-20$ can be
detected, many red dots with small BH masses (simulated QSOs/AGNs) will be below
the black dots (simulated local galaxies), and the $M\bh$-$\sigma$ relation in
QSOs/AGNs will be easily distinguished from the local $M\bhpr$-$\sigma$ relation.
However, the $M\bh-\sigma$ relations with truncation $M_B<-23$ in Figures 4c
and 4d are not as easy to distinguish from the local $M\bhpr$-$\sigma$ relation
as those in Figures 4a and 4b, since QSOs/AGNs housing small BHs are missing
from observations. 

In addition, the observational appearance of the $M\bh$-$\sigma$
relation in QSOs/AGNs may also be affected by obscuration. If
QSOs/AGNs at the early evolutionary stage are strongly obscured, as suggested by
\citet{Fabian99} and \citet{King03}, the $M\bh$-$\sigma$ relation in the observed
QSOs/AGNs should be less deviated from the $M\bhpr$-$\sigma$ relation in nearby
galaxies, since the early stage of BH growth is missing from observations.
However, if the obscuration is a purely geometric effect (as
suggested by the unification model of QSOs/AGNs) and the obscured fraction does
not depend on the QSO/AGN luminosity or central BH mass, the observational
appearance of the $M\bh$-$\sigma$ relation will not be affected by obscuration.
 
\section{Discussion on current observational results of QSOs/AGNs}
\label{sec:dis}
\noindent
Some observations of the BH mass/nuclear luminosity versus velocity dispersion
relation in QSOs/AGNs have been made in the past several years
\citep[e.g.,][]{laor98,wandel99,Geb00b,Ferrarese01,Shields03}. For example,
\citet{Geb00b}
and \citet{Ferrarese01} argue that QSOs/AGNs follow the same BH mass versus
velocity dispersion relation as local galaxies based on some galaxy samples
including a small number (seven or six) of QSOs/AGNs that have measurements of
both BH masses and velocity dispersions. The masses of BHs in QSOs/AGNs in
these studies are estimated by the reverberation mapping technique
\citep[e.g.,][]{WPM99,Kaspi00}.  However, \citet{Krolik01} argues that the BH
mass measured by this method may be either underestimated or overestimated by a
systematic error of up to a factor of $\sim 3$, partly because of the deviation of the
simple assumptions on the dynamics/kinematics and the geometry of broad-line
regions from the reality. For another example, \citet{Shields03} use a larger
sample of QSOs/AGNs
($\sim 100$, including $\sim 20$ high-redshift [$z\ga 1$] QSOs) with estimated
BH mass and velocity dispersion and also argue that the BH mass versus velocity
dispersion relation in QSOs/AGNs is consistent with the relation in nearby
galaxies. In \citet{Shields03}, the BH mass is estimated through the empirical
law between BH mass and broad-line region size \citep[e.g.,][]{Kaspi00}
and the [OIII] line width is used as a surrogate for the galactic velocity
dispersion \citep[e.g.,][]{Nelson00}. These methods might introduce further
uncertainties in BH mass and velocity dispersion. Despite the
measurement
error (or possible systematic error) in BH mass and velocity dispersions, we
note that the flux limit of telescopes can still cause consistency between the
relations in \citet{Shields03}, especially for the subsample of high-redshift
QSOs (which include only BHs with masses higher than $10^9\msun$). Just as
illustrated in Figures~\ref{fig:f3} and ~\ref{fig:f4}, since low-mass
progenitors of big BHs are not detected, the relation in QSOs/AGNs is difficult
to distinguish from that in local galaxies.

Ignoring possible (systematic) measurement errors in BH masses and velocity
dispersions or selection effects for the samples of QSOs/AGNs in \citet{Geb00b}
and \citet{Ferrarese01}, we give a tentative discussion below on the possible
implications of their results.  The
BH masses in QSOs/AGNs have a negative offset of $-0.21\dex$ from local BHs for
the sample in \citet{Geb00b} (including seven QSOs/AGNs) but have a positive
offset for the sample in \citet{Ferrarese01} (including six QSOs/AGNs).
According to equation (\ref{eq:meandiff}), a positive offset might mean that
the velocity dispersion increases during or after the nuclear activity.
However, here we think that the difference of the offsets for these two
different samples is partly because (1) the slope of the local $M\bhpr$-$\sigma$
relation (see $\gamma$ in eq.~\ref{eq:msigma}) in \citet{Ferrarese01} is
steeper than that in \citet{Geb00b}, and (2) the velocity dispersions of two AGNs
(NGC 4051 and NGC 4151, which are included in both samples) measured in
\citet{Ferrarese01} are smaller than those quoted in \citet{Geb00b} by
$\sim10\%$.  According to \citet{Geb00b}, the offset of the $M\bh-\sigma$
relation in QSOs/AGNs from the $M\bhpr-\sigma$ relation in nearby galaxies can
be down to $-0.21$dex. The amount of the offset, based on the calculation
results illustrated in Figure~\ref{fig:f1}(a) and (b), suggests that the period
in which QSOs/AGNs accrete material with Eddington luminosity can be as long
as $\sim1$--$2\tau\Sp$ and the BH mass can be increased by a factor of up to
$\sim$3--8 [i.e., $\sim\exp(1)$--$\exp(2)$] during the nuclear activity (this
suggestion does not contradict the result obtained
in \citealt{YT02} and \citealt{AR02} that local BH mass comes mainly from 
accretion during QSO phases).

Since current samples of QSOs/AGNs with BH mass and velocity dispersion
measurements suffer from the uncertainty of small
number statistics, large measurement errors, and incompleteness due to
the flux limit of telescopes,
a large complete sample of QSOs/AGNs with accurate BH mass and velocity
dispersion measurements is needed to obtain quantitative constraints on the BH
accretion history from the comparison of the BH mass versus velocity
dispersion relation in QSOs/AGNs with that in local galaxies.

\section{Conclusions}\label{sec:con}
\noindent
In this paper we have investigated how the BH mass/nuclear luminosity versus
velocity dispersion relation in QSOs/AGNs is connected with
the local BH mass versus velocity dispersion relation through the mass
growth/nuclear luminosity evolution of individual BHs
(see eqs.~\ref{eq:calPMsigma} and \ref{eq:calPLsigma1}),
by ignoring BH mergers and assuming
that the velocity dispersion does not significantly change during and after the
nuclear activity phase. Circumventing the uncertain history of
triggering the accretion onto seed BHs, the relation in QSOs/AGNs considered in
this paper is for a sample of QSOs/AGNs at all redshifts, rather than for
QSOs/AGNs at a given cosmic time.

Using the observed local BH mass versus velocity dispersion relation
and the assumed form of the QSO/AGN luminosity evolution and BH growth,
we simulate the observational appearance of the BH mass versus velocity dispersion
relation in QSOs/AGNs. 
The simulation results quantitatively show how the BH accretion history (e.g.,
characterized by the timescales $\tau\I$ and $\tau\D$ in this paper;
see \S~\ref{sec:calLM})
affects the difference between the relation in QSOs/AGNs and that in
local galaxies.
A simple example to illustrate this is that: if the BH
mass increases by a factor of $\ga 10$, mainly via Eddington accretion, the
relation in QSOs/AGNs will significantly deviate from the relation in nearby
galaxies, with the mean logarithm of BH mass at a given velocity dispersion
being smaller than that of local BHs by an absolute difference of $\ga 0.3$~\dex
(see Figs.~\ref{fig:f1}a and 1b).
We quantitatively show that if QSOs/AGNs are not always accreting material at
the Eddington rate (e.g., accreting at super-Eddington rates at their
early evolutionary stages or at sub-Eddington rates at their late
evolutionary stages), the distribution function of BH mass at a given velocity
dispersion in QSOs/AGNs can be skewed (i.e., the probability of observing
QSOs/AGNs with low BH mass will be relatively decreased compared to that for
QSOs/AGNs with high BH mass).
We also show that the observational difference between the relation in QSOs/AGNs
and that in local galaxies may be weakened by the selection effect due to
the flux limit of telescopes, since QSOs/AGNs housing small BHs may be missing
from observations. To further
constrain the BH accretion history from the BH mass versus
velocity dispersion relation in QSOs/AGNs, a large, complete QSO/AGN
sample with accurate BH mass and velocity dispersion measurements from
observations is needed.

In addition, the observational appearance of the nuclear luminosity versus velocity
dispersion relation in QSOs/AGNs can also be predicted by the method described
in this paper, and it would also be useful to compare the expectation with future
observations.

We are grateful to the referee for a careful reading of the manuscript and
helpful comments.
Q.Y.  acknowledges support provided by NASA through Hubble Fellowship grant
HF-01169.01-A awarded by the Space Telescope Science Institute, which is
operated by the Association of Universities for Research in Astronomy, Inc.,
for NASA, under contract NAS-5-26555.

\end{document}